\documentclass[12pt]{article}
\usepackage{amsfonts}
\usepackage{amssymb}
\usepackage{xcolor}
\usepackage{color}
\usepackage{graphics}
\usepackage{graphicx}
\begin{document}
\vspace{-48pt}
\title{{\bf Energy of unstable states\\ at long times\footnote{Talk given at \textbf{14th Lomonosov Conference
on Elementary Particle Physics,} Moscow, Russia, August 19 -- 25, 2009.}}}
\author{K. Urbanowski\footnote {e--mail:
K.Urbanowski@proton.if.uz.zgora.pl; K.Urbanowski@if.uz.zgora.pl}, J. Piskorski\footnote{e--mail:
J.Piskorski@proton.if.uz.zgora.pl}\\
University of Zielona G\'{o}ra, Institute of Physics, \\
ul. Prof. Z. Szafrana 4a, 65--516 Zielona G\'{o}ra, Poland.\\}
\date{}

\maketitle
\begin{abstract}
An effect generated by the nonexponential behavior of the survival amplitude of an unstable state
in the long time region is considered. In 1957 Khalfin proved that this  amplitude tends to zero
as $t\rightarrow\infty$  more slowly than any exponential function of $t$. For a time-dependent
decay rate $\gamma(t)$ Khalfin's result means that this $\gamma(t)$ is not a constant for large $t$
but that it tends to zero as $t\rightarrow\infty$. We find that a similar conclusion can be drawn
for the instantaneous energy of the unstable state for a large class of models of unstable states:
This energy tends to the minimal energy of the system ${\cal E}_{min}$ as $t\rightarrow\infty$ which
is much smaller than the energy of this state for $t$ of the order of the lifetime of the considered
state. Analyzing the transition time region between exponential and non-exponential form of the
survival amplitude we find that the instantaneous energy of a considered unstable state can take
large values, much larger than the energy of this state for $t$ from the exponential time region.
Taking into account results obtained for a model considered, it is hypothesized that this purely
quantum mechanical effect may be responsible for the properties of broad resonances such as $\sigma$
meson  as well as having astrophysical and cosmological consequences.
\end{abstract}

\pagebreak

\section{Introduction}

Within the quantum theory the state vector at time $t$, $|\Phi
(t)\rangle$, for the physical system under consideration which
initially (at $t = t_{0} =0$) was in the state $|\Phi\rangle$   can
be found  by solving the  Sch\"{o}dinger equation
\begin{equation}
i\hbar \frac{\partial}{\partial t} |\Phi (t) \rangle = H |\Phi
(t)\rangle, \;\;\;\;\; |\Phi (0) \rangle = |\Phi\rangle, \label
{Schrod}
\end{equation}
where $|\Phi (t) \rangle, |\Phi \rangle \in {\cal H}$,  ${\cal H}$
is the Hilbert space of states of the considered system, $\| \,|\Phi
(t) \rangle \| = \| \,|\Phi \rangle \| = 1$ and $H$ denotes the
total selfadjoint Hamiltonian for the system. If one considers an
unstable state $|\Phi \rangle \equiv |\phi\rangle$ of the system
then using the solution $|\phi (t)\rangle$ of Eq. (\ref{Schrod}) for
the initial condition $|\phi (0) \rangle = |\phi\rangle$ one can
determine the decay law, ${\cal P}_{\phi}(t)$ of this state decaying
in vacuum
\begin{equation}
{\cal P}_{\phi}(t) = |a(t)|^{2}, \label{P(t)}
\end{equation}
where $a(t)$ is  probability amplitude of finding the system at the
time $t$ in the initial state $|\phi\rangle$ prepared at time $t_{0}
= 0$,
\begin{equation}
a(t) = \langle \phi|\phi (t) \rangle . \label{a(t)}
\end{equation}
We have
\begin{equation}
a(0) = 1. \label{a(0)}
\end{equation}
\noindent From basic principles of quantum theory it is known that the
amplitude $a(t)$, and thus the decay law ${\cal P}_{\phi}(t)$ of the
unstable state $|\phi\rangle$, are completely determined by the
density of the energy distribution $\omega({\cal E})$ for the system
in this state \cite{Fock}
\begin{equation}
a(t) = \int_{Spec.(H)} \omega({\cal E})\;
e^{\textstyle{-\frac{i}{\hbar}\,{\cal E}\,t}}\,d{\cal E}.
\label{a-spec}
\end{equation}
where $\omega({\cal E}) \geq 0$.

Note that (\ref{a-spec}) and (\ref{a(0)}) mean that there must be
\begin{equation}
a(0) = \int_{Spec. (H)} \omega ({\cal E})\,d{\cal E} = 1.
\label{a(0)-spec}
\end{equation}
From the last property  and from the Riemann--Lebesgue Lemma it
follows that the amplitude $a(t)$, being the Fourier transform of
$\omega ({\cal E})$ (see (\ref{a-spec})),  must tend to zero as $t
\rightarrow \infty$ \cite{Fock}.

In \cite{Khalfin}
assuming that the spectrum of $H$ must be bounded
from below, $(Spec.(H)\; > \; -\infty)$, and using the Paley--Wiener
Theorem  \cite{Paley}
it was proved that in the case of unstable
states there must be
\begin{equation}
|a(t)| \; \geq \; A\,e^{\textstyle - b \,t^{q}}, \label{|a(t)|-as}
\end{equation}
for $|t| \rightarrow \infty$. Here $A > 0,\,b> 0$ and $ 0 < q < 1$.
This means that the decay law ${\cal P}_{\phi}(t)$ of unstable
states decaying in the vacuum, (\ref{P(t)}), can not be described by
an exponential function of time $t$ if time $t$ is suitably long, $t
\rightarrow \infty$, and that for these lengths of time ${\cal
P}_{\phi}(t)$ tends to zero as $t \rightarrow \infty$  more slowly
than any exponential function of $t$. The analysis of the models of
the decay processes shows that ${\cal P}_{\phi}(t) \simeq
\exp(- \frac{\gamma_{\phi}^{0} t}{\hbar})$, (where
$\gamma_{\phi}^{0}$ is the decay rate of the state $|\phi \rangle$),
to  a very high accuracy  for a wide time range
$t$: From $t$ suitably later than some $T_{0} \simeq t_{0}= 0$ but $T_{0} >
t_{0}$ up to $t \gg \tau_{\phi} = \frac{\hbar}{\gamma_{\phi}^{0}}$
and smaller than $t = t_{as}$, where $t_{as}$ denotes the
time $t$ for which the nonexponential deviations of $a(t)$
begin to dominate (see eg., \cite{Khalfin}, \cite{Goldberger} -- \cite{Greenland}.
From this analysis it follows that in the general
case the decay law ${\cal P}_{\phi}(t)$ takes the inverse
power--like form $t^{- \lambda}$, (where $\lambda
> 0$), for suitably large $t \geq t_{as}\gg \tau_{\phi}$.
 This effect is in
agreement with the  general result (\ref{|a(t)|-as}).  Effects of this type
are sometimes called the "Khalfin effect" \cite{Arbo}.

The problem of  how to detect possible deviations from the exponential form of
${\cal P}_{\phi}(t)$ at the long time region has been attracting the  attention of physicists since the
first theoretical predictions of such an effect. \cite{Greenland,Wessner,Norman1}
Many tests of the decay law performed some time ago
did not indicate
any deviations
from the exponential form of ${\cal P}_{\phi}(t)$ at  the
long time region.
Nevertheless, conditions leading to the nonexponetial behavior
of the amplitude $a(t)$ at long times were studied theoretically \cite{seke} -- \cite{jiitoh}.
Conclusions following from these studies were applied successfully in an experiment described  in the Rothe paper
\cite{rothe},
where the experimental evidence of deviations from the the exponential decay law at long times was
reported. This result gives rise to another problem which now becomes important:
If (and how) deviations from the
exponential decay law at long times affect the energy of the unstable state
and its decay rate at this time region.

Note that in fact the amplitude $a(t)$ contains information about
the decay law ${\cal P}_{\phi}(t)$ of the state $|\phi\rangle$, that
is about the decay rate $\gamma_{\phi}^{0}$ of this state, as well
as the energy ${\cal E}_{\phi}^{0}$ of the system in this state.
This information can be extracted from $a(t)$. Indeed, if
$|\phi\rangle$ is an unstable (a quasi--stationary) state then
\begin{equation}
a(t)  \cong e^{\textstyle{ - \frac{i}{\hbar}({\cal E}_{\phi}^{0} -
\frac{i}{2} \gamma_{\phi}^{0})\,t }}. \label{a-q-stat}
\end{equation}
So, there is
\begin{equation}
{\cal E}_{\phi}^{0} - \frac{i}{2} \gamma_{\phi}^{0} \equiv i
\hbar\,\frac{\partial a(t)}{\partial t} \; \frac{1}{a(t)},
\label{E-iG}
\end{equation}
in the case of quasi--stationary states.

The standard interpretation and understanding of the quantum theory
and the related construction of our measuring devices are such that
detecting the energy ${\cal E}_{\phi}^{0}$ and decay rate
$\gamma_{\phi}^{0}$ one is sure that the amplitude $a(t)$ has the
form (\ref{a-q-stat}) and thus that the relation (\ref{E-iG})
occurs. Taking the above into account one can define the "effective
Hamiltonian", $h_{\phi}$, for the one--dimensional subspace of
states ${\cal H}_{||}$ spanned by the normalized vector
$|\phi\rangle$ as follows  (see, eg. \cite{PRA}),
\begin{equation}
h_{\phi} \stackrel{\rm def}{=}  i \hbar\, \frac{\partial
a(t)}{\partial t} \; \frac{1}{a(t)}. \label{h}
\end{equation}
In general, $h_{\phi}$ can depend on time $t$, $h_{\phi}\equiv
h_{\phi}(t)$. One meets this effective Hamiltonian when one starts
with the Schr\"{o}dinger Equation (\ref{Schrod}) for the total state
space ${\cal H}$ and looks for the rigorous evolution equation for
the distinguished subspace of states ${\cal H}_{||} \subset {\cal
H}$. In the case of one--dimensional ${\cal H}_{||}$  this rigorous
Schr\"{o}dinger--like evolution equation has the following form for
the initial condition $a(0) = 1$ (see, \cite{PRA} and references one finds therein),

\begin{equation}
i \hbar\, \frac{\partial a(t)}{\partial t} \;=\; h_{\phi}(t)\;a(t).
\label{eq-for-h}
\end{equation}
Relations (\ref{h}) and (\ref{eq-for-h}) establish a direct
connection between the amplitude $a(t)$ for the state $|\phi
\rangle$ and the exact effective Hamiltonian $h_{\phi}(t)$ governing
the time evolution in the one--dimensional subspace ${\cal H}_{\|}
\ni |\phi\rangle$. Thus, the use of the evolution equation
(\ref{eq-for-h}) or the relation (\ref{h}) is one of the most
effective tools for the accurate analysis of the early-- as well as
the long--time properties of the energy and decay rate of a given
quasi--stationary state $|\phi (t) \rangle$. 

So let us assume that we know the amplitude $a(t)$. Then starting
with this $a(t)$ and using the expression (\ref{h}) one can
calculate the effective Hamiltonian $h_{\phi}(t)$ in a general case
for every $t$. Thus, one finds the following expressions for the
energy and the decay rate of the system in the state $|\phi\rangle$
under considerations, to be more precise for the instantaneous energy and the instantaneous decay rate,
(for details see: \cite{urbanowski-2-2009}),
\begin{eqnarray}
{\cal E}_{\phi}&\equiv& {\cal E}_{\phi}(t) = \Re\,(h_{\phi}(t),
\label{E(t)}\\
\gamma_{\phi} &\equiv& \gamma_{\phi}(t) = -\,2\,\Im\,(h_{\phi}(t),
\label{G(t)}
\end{eqnarray}
where $\Re\,(z)$ and $\Im\,(z)$ denote the real and imaginary parts
of $z$ respectively.

The deviations of the decay law ${\cal
P}_{\phi}(t)$ from the exponential form can be described
equivalently using time-dependent decay rate (\ref{G(t)}). In terms of such
$\gamma_{\phi}(t)$ the Khalfin observation that ${\cal P}_{\phi}(t)$
must tend to zero as $t \rightarrow \infty$ more slowly than any
exponential function means that $\gamma_{\phi}(t) \ll
\gamma_{\phi}^{0}$ for $t \gg t_{as}$ and $\lim_{t \rightarrow
\infty} \,\gamma_{\phi}(t) = 0$.

Using (\ref{h}) and (\ref{E(t)}), (\ref{G(t)}) one can find that
\begin{eqnarray}
{\cal E}_{\phi} (0) &=& \langle \phi |H| \phi \rangle, \\
{\cal E}_{\phi} (t \sim \tau_{\phi}) & \simeq & {\cal E}_{\phi}^{0} \;\;\neq \;\; {\cal E}_{\phi} (0),\\
\gamma_{\phi}(0) &=& 0,\\
\gamma_{\phi}(t \sim \tau_{\phi}) &\simeq & \gamma_{\phi}^{0}.
\end{eqnarray}

The aim of this talk is to discuss the long time behaviour of ${\cal
E}_{\phi}(t)$  using $a(t)$ calculated for the given density
$\omega({\cal E})$. We show that ${\cal E}_{\phi}(t) \rightarrow {\cal E}_{min} > - \infty$
as $t\rightarrow \infty$ for the model considered and that a wide
class of models has similar long time properties: ${{{\cal
E}_{\phi}(t)}\vline}_{\;t \rightarrow \infty} \neq {\cal
E}_{\phi}^{0}$. It seems that, in contrast to the standard Khalfin
effect \cite{Khalfin},
in the case of the quasi--stationary states
belonging to the same class as excited atomic levels,  these long time
properties of the instantaneous energy ${\cal E}_{\phi}(t)$ have a chance to be
detected, eg.,  by analyzing the properties of the high energy cosmic rays  or the spectra of very distant stars.

\section{The model}

Let us assume that \\${Spec. (H)} = [{\cal E}_{min}, \infty)$,
(where, ${\cal E}_{min} > - \infty$), and let us choose
$\omega ({\cal E})$ as follows
\begin{equation}
\omega ({\cal E}) \equiv \omega_{BW}({\cal E}) = \frac{N}{2\pi}\,  \it\Theta ({\cal E} - {\cal E}_{min}) \
\frac{\gamma_{\phi}^{0}}{({\cal E}-{\cal E}_{\phi}^{0})^{2} +
(\frac{\gamma_{\phi}^{0}}{2})^{2}}, \label{omega-BW}
\end{equation}
where $N$ is a normalization constant and
\[
\it\Theta ({\cal E}) \ = \left\{
  \begin{array}{c}
   1 \;\;{\rm for}\;\; {\cal E} \geq 0, \\
   0 \;\; {\rm for}\;\; {\cal E} < 0,\\
  \end{array}
\right.
\]
For such $\omega_{BW}({\cal E})$ using (\ref{a-spec}) one has
\begin{equation}
a(t) = \frac{N}{2\pi}  \int_{{\cal E}_{min}}^{\infty}
 \frac{{\gamma_{\phi}^{0}}}{({\cal E}-{\cal E}_{\phi}^{0})^{2}
+ (\frac{\gamma_{\phi}^{0}}{2})^{2}}\, e^{\textstyle{ -
\frac{i}{\hbar}{\cal E}t}}\,d{\cal E}, \label{a-BW}
\end{equation}
where
\begin{equation}
\frac{1}{N} = \frac{1}{2\pi} \int_{{\cal E}_{min}}^{\infty}
 \frac{\gamma_{\phi}^{0}}{({\cal E}-{\cal E}_{\phi}^{0})^{2}
+ (\frac{\gamma_{\phi}^{0}}{2})^{2}}\, d{\cal E}. \label{N}
\end{equation}
Formula  (\ref{a-BW}) leads to the result
\begin{eqnarray}
a(t) &=& N\,e^{\textstyle{- \frac{i}{\hbar} ({\cal
E}_{\phi}^{0} -
i\frac{\gamma_{\phi}^{0}}{2})t}}\,
\Big\{1 - \frac{i}{2\pi} \times\nonumber \\
&& \times\,\Big[
e^{\textstyle{\frac{\gamma_{\phi}^{0}t}{\hbar}}}\,
E_{1}\Big(-\frac{i}{\hbar}({\cal E}_{\phi}^{0} -{\cal E}_{min}
+ \frac{i}{2} \gamma_{\phi}^{0})t\Big) \nonumber\\
&&\,-\, E_{1}\Big(- \frac{i}{\hbar}({\cal E}_{\phi}^{0} -{\cal E}_{min} -
\frac{i}{2} \gamma_{\phi}^{0})t\Big)\,\Big]\, \Big\}, \label{a-E(1)}
\end{eqnarray}
where $E_{1}(x)$ denotes the integral--exponential function \cite{Sluis,Abramowitz}.

In general one has
\begin{equation}
a(t) \equiv a_{exp}(t) + a_{non}(t),
\label{a-exp+a-non}
\end{equation}
where
\[
a_{exp}(t) = N\,e^{\textstyle{- \frac{i}{\hbar} ({\cal
E}_{\phi}^{0} -
i\frac{\gamma_{\phi}^{0}}{2})t}}, \;\;\;\;\;\;\;\; a_{non}(t) = a(t) - a_{exp}(t).
\]

Making use of  the asymptotic expansion of $E_{1}(x)$ \cite{Abramowitz}
\begin{equation}
{E_{1}(z)\vline}_{\, |z| \rightarrow \infty} \;\;\sim \;\;
\frac{e^{\textstyle{ -z}}}{z}\,( 1 - \frac{1}{z} + \frac{2}{z^{2}} -
\ldots ),  \label{E1-as}
\end{equation}
where $| \arg z  | < \frac{3}{2} \pi$, one finds
\begin{eqnarray}
{a(t)\vline}_{\, t \rightarrow \infty} &\simeq & N
e^{\textstyle - \frac{i}{\hbar}\,h_{\phi}^{0}\,t}
\nonumber \\&&
\;+\;
\frac{N}{2 \pi}\;e^{\textstyle{-\frac{i}{\hbar}\,{\cal E}_{min}t}}\;
\Big\{
(- i)\; \frac{
\gamma_{\phi}^{0}}{|\,h_{\phi}^{0}-{\cal E}_{min}\,|^{\,2}}
 \, \frac{\hbar}{t}  \nonumber \\
&&\,- 2\,\frac{({\cal E}_{\phi}^{0}\,-\,{\cal E}_{min})\,
\gamma_{\phi}^{0}}{|\,h_{\phi}^{0}\,-\,{\cal E}_{min}\,|^{\,4}} \,
\Big(\frac{\hbar}{t}\Big)^{2}\,+ \ldots\Big\} \label{a(t)-as}
\end{eqnarray}
where $h^{0}_{\phi} = {\cal E}_{\phi}^{0} \,- \,\frac{i}{2}\,\gamma_{\phi}^{0}$,
and
\begin{eqnarray}
{h_{\phi}(t)\vline}_{\,t \rightarrow \infty} & = &
{i \hbar \,\frac{\partial a(t)}{\partial
t}\,\frac{1}{a(t)} \vline}_{\,t \rightarrow \infty} \nonumber \\
&\simeq & {\cal E}_{min}\,
-\,i\,\frac{\hbar}{t}\;  - \;2\, \frac{ {\cal E}_{\phi}^{0}\,-\,{\cal E}_{min}}{
|\,h_{\phi}^{0}\,-\,{\cal E}_{min} \,|^{\,2} }  \; \Big( \frac{\hbar}{t} \Big)^{2}
\;+\ldots \;\; \label{h-as}
\end{eqnarray}
for the considered case (\ref{omega-BW}) of $\omega_{BW}({\cal E})$
(for details see \cite{urbanowski-2-2009}).
From (\ref{h-as}) it follows that
\begin{eqnarray}
\Re\,({h_{\phi}(t)\vline}_{\,t \rightarrow \infty}) &\stackrel{\rm
def}{=}& {\cal E}_{\phi}^{\infty} (t)\, \nonumber \\
&\simeq & {\cal E}_{min}
 -\,2\,
\frac{ {\cal E}_{\phi}^{0}\,-\,{\cal E}_{min}}{ |\,h_{\phi}^{0}\,-\,{\cal E}_{min} \,|^{\,2} }  \; \Big(
\frac{\hbar}{t} \Big)^{2} \nonumber \\ &&
                            \begin{array}{c}
                               {} \\
                               \longrightarrow \\
                               \scriptstyle{t \rightarrow \infty}
                             \end{array}
                             \, {\cal E}_{min},\label{Re-h-as}
\end{eqnarray}
where ${\cal E}_{\phi}^{\infty}(t) = {\cal
E}_{\phi}(t)|_{\,t \rightarrow \infty}$,
and
\begin{equation}
\Im\,({h_{\phi}(t)\vline}_{\,t \rightarrow \infty}) \simeq
-\,\frac{\hbar}{t} \,
\begin{array}{c}
                               {} \\
                               \longrightarrow \\
                               \scriptstyle{t \rightarrow \infty}
                             \end{array}
\,0. \label{Im-h-as}
\end{equation}
The property (\ref{Re-h-as}) means that
\begin{equation}
\Re\,({h_{\phi}(t)\vline}_{\,t \rightarrow \infty})\,\equiv\,
{\cal E}_{\phi}^{\infty}(t)\,
<\, {\cal E}_{\phi}^{0}. \label{E-infty<E0}
\end{equation}
For different states $|\phi\rangle \, =\,|j\rangle$, ($j = 1,2,3,\ldots$)
one has
\begin{equation}
\Im\,({h_{1}(t)\vline}_{\,t \rightarrow \infty}) =
\Im\,({h_{2}(t)\vline}_{\,t \rightarrow
\infty}),\label{Im-h1-h2-as}
\end{equation}
whereas in general $\gamma_{1}^{0}\, \neq \,\gamma_{2}^{0}$.

Note that
from (\ref{a(t)-as}) one obtains
\begin{eqnarray}
{\vline\,{a(t)\vline}_{\, t \rightarrow
\infty}\,\vline}^{\,2} &\simeq& N^{2} e^{\textstyle - \frac{
\gamma_{\phi}^{0}}{\hbar}\,\,t} \nonumber \\
&& + \frac{N^{2}}{\pi}\,\,\,\sin\,[({\cal E}_{\phi}^{0} - {\cal E}_{min})\,t]\,\,\,\,\,
e^{\textstyle - \frac{1}{2}\,\frac{\gamma_{\phi}^{0}}{\hbar}\,\,t}\;\;
\frac{\gamma_{\phi}^{0}}{|h_{\phi}^{0}\,-\,{\cal E}_{min}\,|^{\,2}} \;
\frac{\hbar}{t} \nonumber \\
&& + \frac{N^{2}}{4
\pi^{2}}\;\frac{(\gamma_{\phi}^{0})^{2}}{|h_{\phi}^{0}\,-\,{\cal E}_{min}\,|^{\,4}} \;
\frac{\hbar^{2}}{t^{2}}\; + \;\ldots\;\; . \label{t-as-1}
\end{eqnarray}
Relations (\ref{a(t)-as}) ---  (\ref{Im-h1-h2-as})
become important
for times $t > t_{as}$, where $t_{as}$ denotes the time $t$ at which
contributions to ${\vline\,{a(t)\vline}_{\, t \rightarrow
\infty}\,\vline}^{\,2}$ from the first exponential component in
(\ref{t-as-1}) and from the third component proportional to
$\frac{1}{t^{2}}$ are comparable, that is (see (\ref{a-exp+a-non})),
\begin{equation}
|a_{exp}(t)|^{2} \,\simeq\,|a_{non}(t)|^{2}
\label{a-exp=a-non}
\end{equation}
for $t\rightarrow \infty $.
So $t_{as}$ can be be found by
considering the following relation
\begin{equation}
e^{\textstyle - \frac{ \gamma_{\phi}^{0}}{\hbar}\,\,t}\; \sim\;
\frac{1}{4
\pi^{2}}\;\frac{(\gamma_{\phi}^{0})^{2}}{|h_{\phi}^{0}\,-\,{\cal E}_{min}\,|^{\,4}} \;
\frac{\hbar^{2}}{t^{2}}. \label{t-as-2}
\end{equation}
Assuming that the right hand side is equal to the left hand side in
the above relation one gets a transcendental equation.  Exact
solutions of such an equation can be expressed by means of the
Lambert $W$ function \cite{Corless}.
An asymptotic solution of the
equation obtained from the relation (\ref{t-as-2}) is relatively
easy to find \cite{Olver}.
The very approximate asymptotic solution,
$t_{as}$, of this equation for $(\frac{{\cal
E}_{\phi}}{\gamma_{\phi}^{0}})\,>\,10$ (in general for
$(\frac{{\cal E}_{\phi}}{\gamma_{\phi}^{0}})\,\rightarrow \,\infty$)
has the form
\begin{eqnarray}
\frac{\gamma_{\phi}^{0}\,t_{as}}{\hbar} &\simeq & 8,28 \,+\, 4\,
\ln\,(\frac{{\cal E}_{\phi}^{0}\,-\,{\cal E}_{min}}{\gamma_{\phi}^{0}}) \nonumber \\
&&+\, 2\,\ln\,[8,28 \,+\,4\,\ln\,(\frac{{\cal
E}_{\phi}^{0}\,-\,{\cal E}_{min}}{\gamma_{\phi}^{0}})\,]\,+\, \ldots \;\;.
\label{t-as-3}
\end{eqnarray}

\section{Some generalizations}

To complete the analysis performed in the previous Section let us
consider a more general case of $\omega ({\cal E})$ and $a(t)$.
For a start, let us consider a   relatively simple case when $\lim_{{\cal E} \rightarrow {\cal E}_{min}+}
\;\omega ({\cal E})\stackrel{\rm def}{=} \omega_{0}>0$ and ${\omega ({\cal E})\vline}_{\;{\cal E}\, < \,{\cal E}_{min}}\,=\,0$.   Let  derivatives  $\omega^{(k)}({\cal E})$, ($k= 0,1,2, \ldots, n$),  be continuous
in 
$[ {\cal E}_{min}, \infty)$, (that is let for ${\cal E} > {\cal E}_{min}$ all
$\omega^{(k)}({\cal E})$ be continuous
and all the limits 
$\lim_{{\cal E} \rightarrow {\cal E}_{min}+}\,\omega^{(k)}({\cal E})$ exist)  and
let all these $\omega^{(k)}({\cal E})$ be absolutely integrable functions then
(see \cite{urbanowski-1-2009}),
\begin{equation}
a(t) \; \begin{array}{c}
          {} \\
          \sim \\
          \scriptstyle{t \rightarrow \infty}
        \end{array}
        \;- \frac{i\hbar}{t}\;e^{\textstyle{-\frac{i}{\hbar}{\cal E}_{min} t}}\;
        \sum_{k = 0}^{n-1}(-1)^{k} \,\big(\frac{i\hbar}{t}\big)^{k}\,\omega^{(k)}_{0},
        \label{a-omega}
\end{equation}
where $\omega^{(k)}_{0}  \stackrel{\rm def}{=} \lim_{{\cal E}\rightarrow {\cal E}_{min}+}
\;\omega^{(k)} ({\cal E})$.

Let us now consider a more complicated form of the density $\omega ({\cal E})$. Namely
let $\omega ({\cal E})$ be of the form
\begin{equation}
\omega ({\cal E}) = ( {\cal E} - {\cal E}_{min})^{\lambda}\;\eta ({\cal E})\; \in \; L_{1}(-\infty, \infty),
\label{omega-eta}
\end{equation}
where $0 < \lambda < 1$ and it is assumed that $\eta^{(k)}({\cal E})$,
($k= 0,1,\ldots, n$), \linebreak exist and they are continuous
in $[{\cal E}_{min}, \infty)$, and  limits
$\lim_{{\cal E} \rightarrow {\cal E}_{min}+}\;\eta^{(k)}({\cal E})$ exist,
$\lim_{{\cal E} \rightarrow \infty}\;( {\cal E} - {\cal E}_{min})^{\lambda}\,\eta^{(k)}({\cal E}) = 0$
for all above mentioned $k$ and \linebreak ${\omega ({\cal E})\vline}_{\;{\cal E}\, < \,{\cal E}_{min}}\,=\,0$, then
\begin{eqnarray}
a(t) & \begin{array}{c}
          {} \\
          \sim \\
          \scriptstyle{t \rightarrow \infty}
        \end{array} &
        - \frac{i\hbar}{t}\;\lambda\;e^{\textstyle{-\frac{i}{\hbar}{\cal E}_{min} t}}\;
        \Big[\alpha_{n}(t) + \big(-\,\frac{i\hbar}{t}\,\big)\,\alpha_{n-1}(t) \nonumber \\
        && + \big(-\,\frac{i\hbar}{t}\big)^{2}\,\alpha_{n-2}(t)\nonumber \\&& +
        \big(-\,\frac{i\hbar}{t}\big)^{3}\,\alpha_{n-3}(t)+ \dots \Big],
        \label{a-eta}
\end{eqnarray}
where (compare \cite{erdelyi,copson})
\begin{equation}
\alpha_{n-k}(t) = \sum_{l=0}^{n-k-1}\,\frac{\Gamma (l + \lambda)}{l!}\;\,e^{\textstyle{-\,i\,\frac{\pi (l + \lambda +2)}{2}}}
\;\eta_{0}^{(l + k)}\,\big(\frac{\hbar}{t} \big)^{l + \lambda}, \label{alpha}
\end{equation}
$\Gamma (z)$ is the Gamma Function and $\eta^{(j)}_{0}  = \lim_{{\cal E} \rightarrow {\cal E}_{min}+}
\;\eta^{(j)} ({\cal E})$, $\eta^{(0)}({\cal E}) = \eta ({\cal E})$ and $j = 0,1, \ldots,n$.

The asymptotic form of $h_{u}(t)$ for $t \rightarrow \infty$ for the $a(t)$ given by the relation  (\ref{a-omega}) looks as follows
\begin{eqnarray}
h_{u}^{\infty}(t)\; \stackrel{\rm def}{=}\;
         {h_{u}(t)\,\vline}_{\;t \rightarrow \infty} \; &=& \;{\cal E}_{min} \,-\,i\,\frac{\hbar}{t}\nonumber \\
         &&-\frac{\omega_{0}^{(1)}}{\omega_{0}}\;\big(\,\frac{\hbar}{t}\,\big)^{2}\;+\;\ldots\;\, . \label{h-infty}
\end{eqnarray}

In the more general case  of $a(t)$ (see, e.g. (\ref{a-eta}) ) after some algebra the
asymptotic approximation  of $a(t)$  can be written as follows
\begin{equation}
a(t) \;\;
\begin{array}{c}
   {} \\
   \sim\\
   {\scriptstyle t \rightarrow \infty}
 \end{array}
\;\;e^{\textstyle{-i\frac{t}{\hbar}\,{\cal E}_{min}}}\; \sum_{k=0}^{N} \,\frac{c_{k}}{t^{\xi + k}},
\label{a(t)-as-w}
\end{equation}
where $\xi \geq 0$ and $c_{k}$ are complex numbers.

From the relation (\ref{a(t)-as-w})
one concludes that
\begin{equation}
\frac{\partial a(t)}{\partial t} \;\;
\begin{array}{c}
   {} \\
   \sim\\
   {\scriptstyle t \rightarrow \infty}
 \end{array}
\;\; e^{\textstyle{-i\frac{t}{\hbar}\,{\cal E}_{min}}}\;\Big\{-\frac{i}{\hbar}\, {\cal E}_{min}\,
-\,\sum_{k=0}^{N} \,(\xi + k)\,\frac{c_{k}}{t^{\xi + k
+ 1}}\Big\}. \label{da(t)-as-w}
\end{equation}

Now let us take into account the relation (\ref{eq-for-h}). From
this relation and relations (\ref{a(t)-as-w}), (\ref{da(t)-as-w}) it
follows that
\begin{equation}
h_{\phi}(t) \;\;
\begin{array}{c}
   {} \\
   \sim\\
   {\scriptstyle t \rightarrow \infty}
 \end{array}
\;\;
{\cal E}_{min}\,+\,
\frac{d_{1}}{t}\, +
\,\frac{d_{2}}{t^{2}}\,+\,\frac{d_{3}}{t^{3}}\,+\,\ldots\,\,\, ,
\label{h-sim-w}
\end{equation}
where $d_{1}, d_{2}, d_{3}, \ldots$ are complex numbers with negative or positive real and imaginary parts. This means
that in the case of the asymptotic approximation to $a(t)$ of
the form (\ref{a(t)-as-w}) the following  property holds,
\begin{equation}
\lim_{t \rightarrow \infty}\,h_{\phi}(t) \, = \, {\cal E}_{min}\, <\, {\cal E}_{\phi}^{0}. \label{lim-h}
\end{equation}

It seems to be important that  results (\ref{h-sim-w}) and
(\ref{lim-h}) coincide with the results (\ref{h-as}) ---
(\ref{Im-h1-h2-as})
obtained for the density $\omega_{BW}({\cal E})$
given by the formula (\ref{omega-BW}). This means that general
conclusion obtained for the other $\omega ({\cal E})$ defining
unstable states should be similar to those following from
(\ref{h-as}) --- (\ref{Im-h1-h2-as}).

\section{Final remarks.}
Long time properties of the survival probability $|a(t)|^{2}$ and instantaneous energy ${\cal E}_{\phi}(t)$ are relatively easy to find analytically for times $t \gg t_{as}$ even in the general case as it was shown in previous Section and \cite{urbanowski-1-2009}. It is much more difficult to analyze these properties analytically in the transition time region where $t \sim t_{as}$. It can be done numerically for given models.

The model considered in Sec. 2 and defined by the density $\omega_{BW}({\cal E})$, (\ref{omega-BW}), allows one to find numerically the decay curves and the instantaneous energy $\varepsilon_{\phi}(t)$ as a function of time $t$. The results presented in this Section have been obtained assuming for simplicity that the minimal energy ${\cal E}_{min}$ appearing in the formula (\ref{omega-BW}) is equal to zero,   ${\cal E}_{min} = 0$. So, all numerical calculations were performed for the density $\tilde{\omega}_{BW}({\cal E})$ given by the following formula
\begin{equation}
\tilde{\omega}_{BW}({\cal E}) = \frac{N}{2\pi}\,  \it\Theta ({\cal E}) \
\frac{\gamma_{\phi}^{0}}{({\cal E}-{\cal E}_{\phi}^{0})^{2} +
(\frac{\gamma_{\phi}^{0}}{2})^{2}}, \label{omega-BW-E-min=0}
\end{equation}
for some chosen $\frac{{\cal E}_{\phi}^{0}}{\gamma_{\phi}^{0}}$. Performing calculations  particular attention was paid to the form of the probability $|a(t)|^{2}$, i. e. of the decay curve, and of the instantaneous energy $\varepsilon_{\phi}(t)$ for times $t$ belonging to the most interesting transition time-region between exponential and nonexponential parts of $|a(t)|^{2}$, where the following relation corresponding with (\ref{a-exp=a-non}) and (\ref{t-as-2}) takes place,
\begin{equation}
|a_{exp}(t)|^{2} \;\sim \;  |a_{non}(t)|^{2},
\label{a-exp=a-non-1}
\end{equation}
where $a_{exp}(t), a_{non}(t)$ are defined by (\ref{a-exp+a-non}). Results are presented graphically below.
\begin{figure}
\includegraphics[height=60mm,width=110mm]{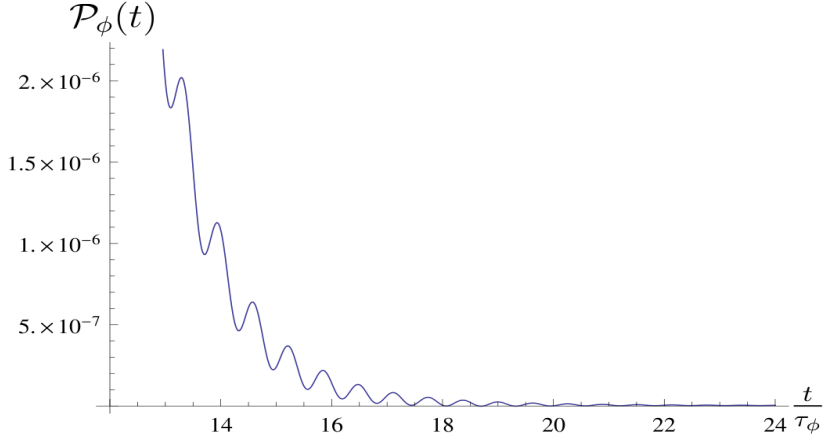}\\
\caption{Survival probability ${\cal P}_{\phi}(t) = |a(t)|^{2}$ in the transition time region.
The case $\frac{{\cal E}_{\phi}^{0}}{\gamma_{\phi}^{0}} = 10$.}
\label{a10-1}
\end{figure}

\begin{figure}
\begin{center}
\includegraphics[height=60mm,width=110mm]{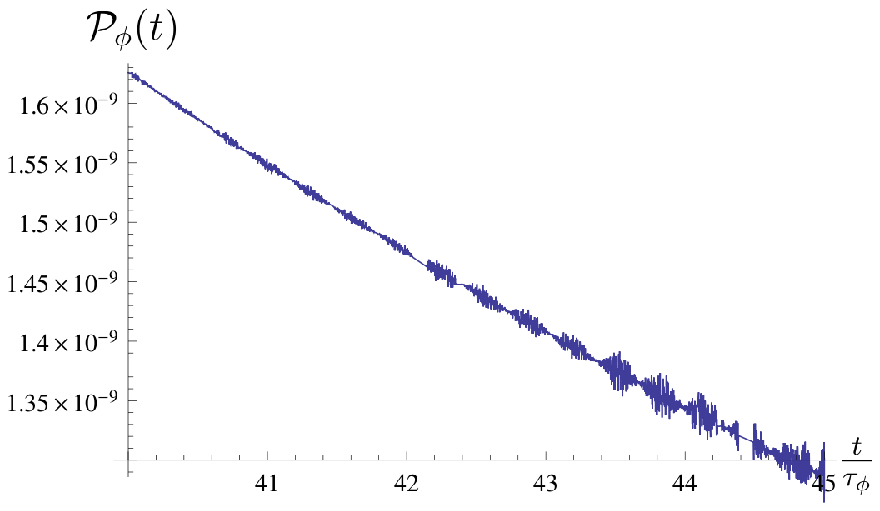}\\
\caption{Survival probability ${\cal P}_{\phi}(t) = |a(t)|^{2}$ in the transition time region.
The case $\frac{{\cal E}_{\phi}^{0}}{\gamma_{\phi}^{0}} = 10$.}
\label{a10-2}
\end{center}
\end{figure}

\begin{figure}
  \includegraphics[width=120mm]{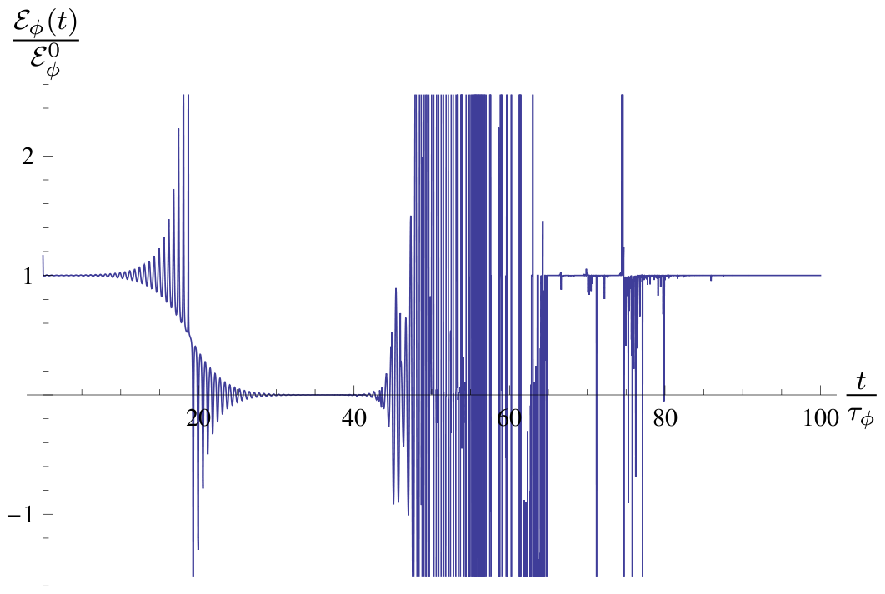}\\
  \caption{Instantaneous energy ${\cal E}_{\phi}(t)$ in the transition time region. The case $\frac{{\cal E}_{\phi}^{0}}{\gamma_{\phi}^{0}} = 10$.}
  \label{h10}
\end{figure}

\begin{figure}
\begin{center}
\includegraphics[height=60mm,width=110mm]{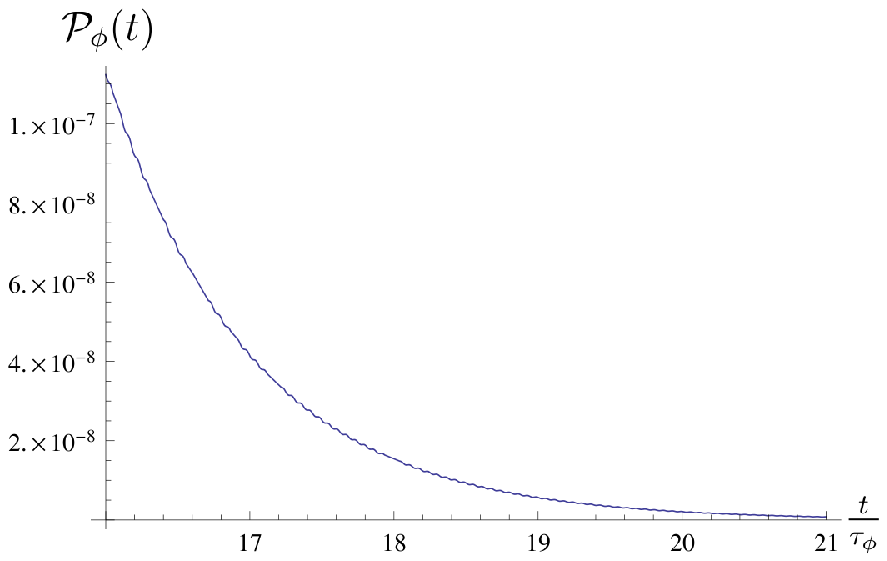}\\
\caption{Survival probability ${\cal P}_{\phi}(t) = |a(t)|^{2}$ in the transition time region.
The case $\frac{{\cal E}_{\phi}^{0}}{\gamma_{\phi}^{0}} = 100$.}
\label{a100-1}
\end{center}
\end{figure}

\begin{figure}
\begin{center}
\includegraphics[height=60mm,width=110mm]{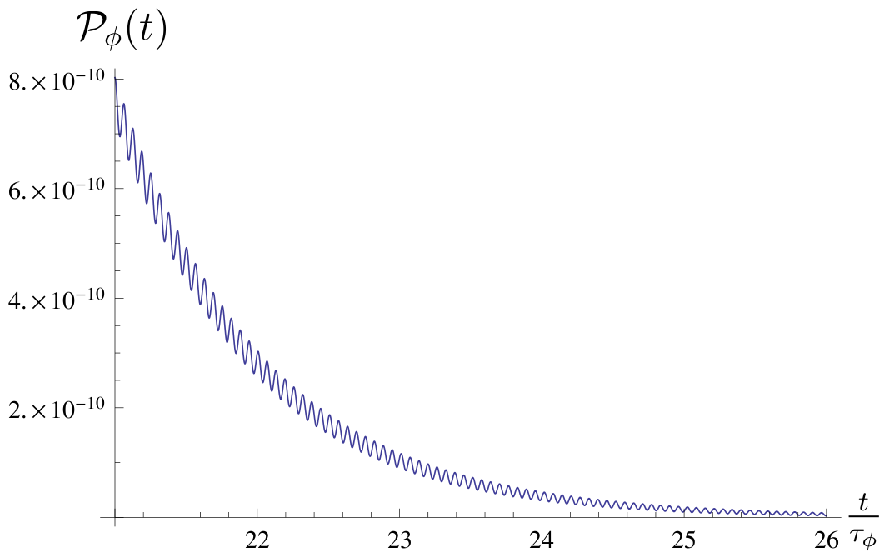}\\
\caption{Survival probability ${\cal P}_{\phi}(t) = |a(t)|^{2}$ in the transition time region.
The case $\frac{{\cal E}_{\phi}^{0}}{\gamma_{\phi}^{0}} = 100$.}
\label{a100-2}
\end{center}
\end{figure}

\begin{figure}
\begin{center}
\includegraphics[height=60mm,width=110mm]{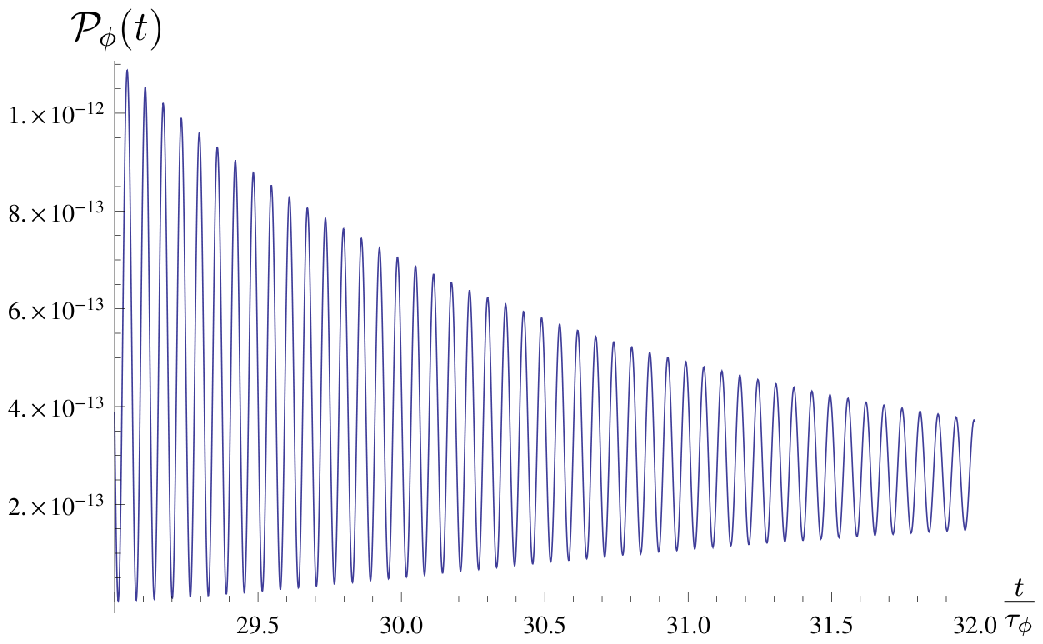}\\
\caption{Survival probability ${\cal P}_{\phi}(t) = |a(t)|^{2}$ in the transition time region.
The case $\frac{{\cal E}_{\phi}^{0}}{\gamma_{\phi}^{0}} = 100$.}
\label{a100-3}
\end{center}
\end{figure}

\begin{figure}
  \includegraphics[width=120mm]{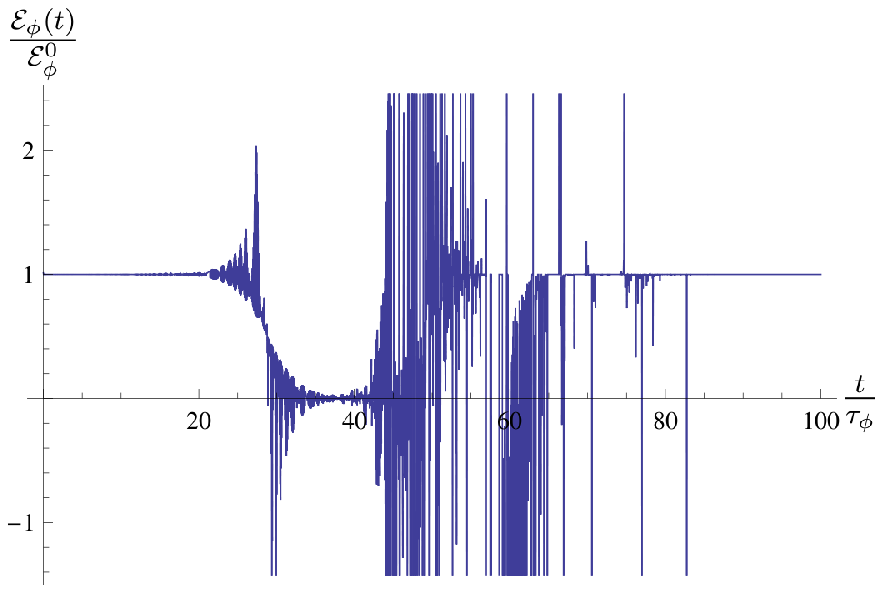}\\
  \caption{Instantaneous energy ${\cal E}_{\phi}(t)$ in the transition time region. The case $\frac{{\cal E}_{\phi}^{0}}{\gamma_{\phi}^{0}} = 100$.}
  \label{h100}
\end{figure}

A similar form of a decay curves one meets for a very large class of models defined  by energy densities $\omega({\cal E })$
of the following type (see \cite{Fonda,nowakowski-3}),
\begin{eqnarray}
\omega({\cal E})\,&=&\, \frac{N}{2\pi}\;{\mathit \Theta}({\cal E} - {\cal E}_{min} )\;({\cal E} - {\cal E}_{min})^{\lambda} \, \frac{\gamma_{\phi}^{0}}{({\cal E}- {\cal E}_{\phi}^{0})^{2}+
\frac{(\gamma_{\phi}^{0})^{2}}{4}}\, f({\cal E}),\label{omega-gen}\\
&\equiv & \omega_{BW}({\cal E})\;({\cal E} - {\cal E}_{min})^{\lambda}\, f({\cal E}), \nonumber
\end{eqnarray}
where $\lambda \geq 0$, $f({\cal E})$ is a form--factor ---  it is a smooth function going to zero as ${\cal E} \rightarrow \infty$ and it has no threshold and no pole. The asymptotical large time behavior of $a(t)$ is due to the term $({\cal E} - {\cal E}_{min})^{\lambda}$ and the choice of $\lambda$ (see Sec. 3). The density $\omega({\cal E })$ defined by the relation (\ref{omega-gen}) fulfills all physical requirements and it leads to the decay curves having a very similar form at transition times region to the decay curves presented above . The characteristic  feature of all these decay curves is the presence of sharp and frequent oscillations at the transition times region (see Figs (\ref{a10-1}), (\ref{a10-2}), (\ref{a100-1}), (\ref{a100-2}), (\ref{a100-3}) ). This means that derivatives of the amplitude $a(t)$ may reach extremely large values for some times from the transition time region and the modulus of these derivatives is much larger than the modulus of $a(t)$, which is very small for these times. This explains why in this time region the real and imaginary parts of $h_{\phi}(t) \equiv {\cal E}_{\phi}(t) \, - \,\frac{i}{2}\,\gamma_{\phi}(t)$, which can be expressed by the relation (\ref{h}), ie. by a large  derivative of $a(t)$  divided by a very small $a(t)$, reach values much larger than the energy ${\cal E}_{\phi}^{0}$ of the the unstable state measured at times for which the decay curve has the exponential form. For the model considered we found that, eg. for $\frac{{\cal E}_{\phi}^{0}}{\gamma_{\phi}^{0}} = 10$ and   $5 \tau_{\phi}\, \leq \,t\,\leq\,60 \tau_{\phi}$  the maximal value of the instantaneous energy equals ${\cal E}_{\phi}(t)\, = \,89,2209 \,{\cal E}_{\phi}^{0}$ and  ${\cal E}_{\phi}(t)$ reaches this value for $t\equiv t_{mx,\;10} = 53,94\,\tau_{\phi}$   and then the survival probability ${\cal P}_{\phi}(t)$ is of order ${\cal P}_{\phi}(t_{mx,\,10}) \sim 10^{-9}$.

The question is whether and where this effect can manifest itself. There are two possibilities to observe the above long time properties of unstable states: The first one is that one should analyze properties of unstable states having not too long values of $t_{as}$. The second one is finding a possibility to observe a suitably large number of events, i.e. unstable particles, created by the same source.

The problem with understanding the properties of broad resonances in the scalar sector ($\sigma$ meson problem \cite{pdg-2008}) discussed in \cite{nowakowski-1,nowakowski-2}, where the hypothesis was formulated that this problem could be connected with properties of the decay amplitude in the transition time region,
seems to be  possible manifestations of this effect and this problem refers  to the first possibility mentioned above. There is the problem with determining  the mass of broad resonances. The measured range of possible mass of $\sigma$ meson is very wide, 400 -- 1200 MeV. So one can not exclude the possibility that the masses of some $\sigma$  mesons  are measured for times of order  their lifetime and some of them for times where their instantaneous energy ${\cal E}_{\sigma}(t)$ is much larger. This is exactly the case presented in Fig. (\ref{h10}) and Fig. (\ref{h100}). For broad mesons the ratio $\frac{{\cal E}_{\sigma}^{0}}{\gamma_{\sigma}^{0}}$ is relatively small and thus the time $t_{as}$ when the above discussed effect occurs appears to be not too long.

Astrophysical and cosmological processes in which  extremely huge numbers of unstable particles are created
seem to be  another possibility for the above discussed effect to become manifest. The probability ${\cal P}_{\phi}(t) = |a(t)|^{2}$ that an unstable particle, say $\phi$, survives up to time $t \sim t_{as}$ is extremely small. Let ${\cal P}_{\phi}(t)$ be
\begin{equation}
{ {\cal P}_{\phi}(t)\,\vline }_{\;t \sim t_{as}}\;\sim\;10^{-k},
\label{p(t)-k}
\end{equation}
where $k \gg 1$, then there is a chance to observe some of particles $\phi$ survived at $t \sim t_{as}$ only if there is a source creating these particles in ${\cal N}_{\phi}$ number such that
\begin{equation}
{{\cal P}_{\phi}(t)\,\vline }_{\;t \sim t_{as}}\;{\cal N}_{\phi} \; \gg \;1.
\label{N-phi}
\end{equation}
So if  a source exists that creates a flux containing
\begin{equation}
{\cal N}_{\phi} \;\sim\;10^{\,l},
\label{N-phi-l}
\end{equation}
unstable particles and $l \gg k$ then the probability theory states that the number $N_{surv}$ unstable particles
\begin{equation}
N_{surv} = { {\cal P}_{\phi}(t)\,\vline }_{\;t \sim t_{as}}\;{\cal N}_{\phi} \;\sim\;10^{l - k} \; \gg\;0,
\label{N-surv}
\end{equation}
has to survive up to time $t \sim t_{as}$. Sources creating such numbers of unstable particles are known from cosmology and astrophysics.
The Big Bang is the obvious example of such a source.  Some other examples include processes taking place in  galactic  nuclei (galactic cores) and inside  stars, etc.

\begin{figure}
\begin{center}
  \includegraphics[width=110mm]{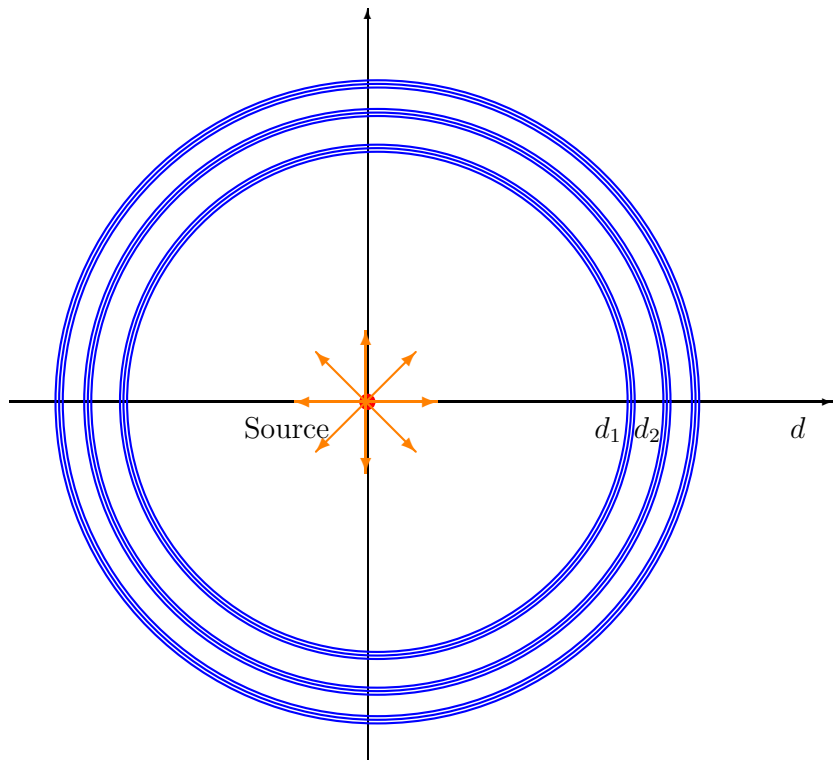}\\
  \end{center}
  \caption{Halos surrounding a source of unstable particles}
  \label{circles-a}
\end{figure}

So let us assume that we have an astrophysical source creating a sufficiently large number of unstable particles in unit of time and emitting a flux of these particles and that this flux is constant or slowly varying in time.
Consider as an example a flux of neutrons. From (\ref{t-as-3}) it follows that for the neutron $t_{as}^{n} \sim \,(250 \tau_{n} - \,300 \tau_{n})$, where $\tau_{n} \simeq 886$ [s]. If the energies of these neutrons are of order $30 \times 10^{17}$ [eV] then
during time $ t  \sim t_{as}^{n}$
they can reach a distance $d_{n} \sim 25 000$ [ly], that is the distance of about a half of  the Milky Way radius. Now if in a unit ot time a suitably large number of neutrons ${\cal N}_{n}$ of the energies mentioned is created by this source then in the distance $d_{n}$ from the source
a number of spherically symmetric space areas (halos) surrounding  the source, where neutron instantaneous energies ${\cal E}_{n}(t)$ are much larger than  their rest energy ${\cal E}_{n}^{0} = \frac{m_{n}^{0} \,c^{2}}{\sqrt{1 - (\frac{v_{n}}{c})^{2}}}$, ($m_{n}^{0}$ is the neutron rest mass and $v_{n}$ denotes its velocity) have to appear (see Fig. (\ref{circles-a})). Of course this conclusion holds also for other unstable particles $\phi_{\alpha}$ produced by this source.

Every kind of  particles $\phi_{\alpha}$ has its own halos located at distances $d_{k}^{\phi_{\alpha}}$,
\[
d_{k}^{\phi_{\alpha}} \sim v_{\phi_{\alpha}}\,t_{as}^{\phi,\,k},\;\;\;\;(k =1,2, \ldots),
\]
from the source. Radiuses  $d_{k}^{\phi_{\alpha}}$ of these halos are determined by the particles'  velocities $v_{\phi_{\alpha}}$ and by times  $t_{as}^{\phi,\,k}$
when instantaneous energies ${\cal E}_{\phi_{\alpha}}(t)$ have local maxima.

\begin{figure}
\begin{center}
  \includegraphics[width=110mm]{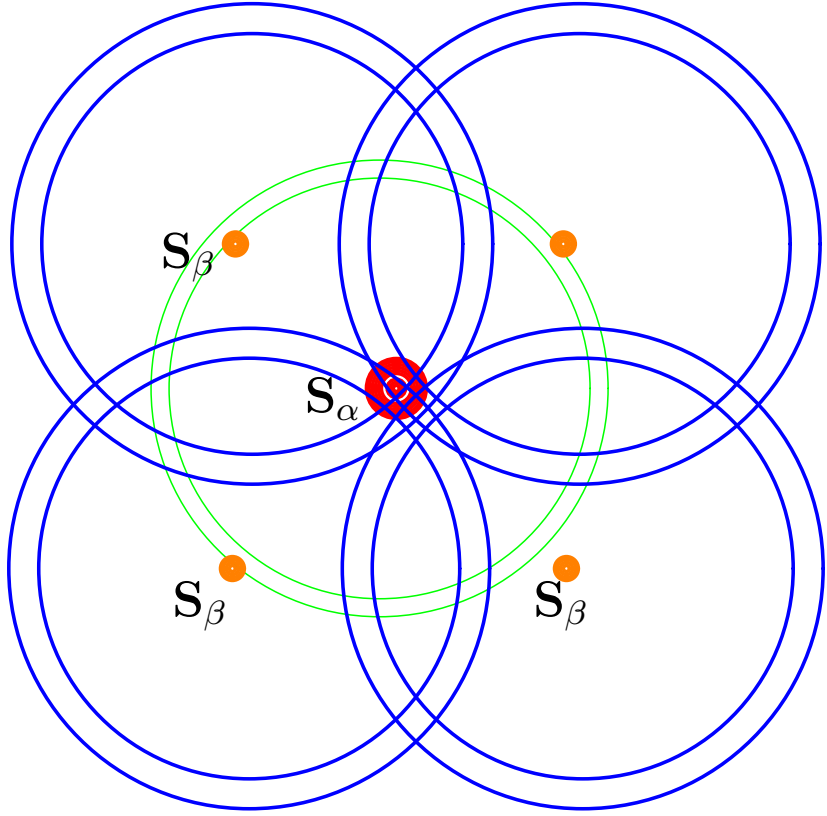}\\
  \end{center}
  \caption{Halos generated by a few not too distant  sources ${\bf S_{\alpha}}, {\bf S_{\beta}}$.}
  \label{circles-b}
\end{figure}

Unstable particles $\phi_{\alpha}$  forming these halos and having instantaneous energies   ${\cal E}_{\phi_{\alpha}}(t) \gg {\cal E}_{\phi_{\alpha}}^{0} = m_{\phi_{\alpha}}^{0}\,c^{2}$ have to interact gravitationally with objects outside of these halos as particles of masses $m_{\phi_{\alpha}}(t) = \frac{1}{c^{2}}\,   {\cal E}_{\phi_{\alpha}}(t) \,\gg\, m_{\phi_{\alpha}}^{0}$. The possible observable effects depend on the astrophysical source of these particles considered.

If the halos are formed by unstable particles emitted as a result of internal star processes then
in the case of very young stars cosmic dust and gases should be attracted by these halos as a result of a gravity attraction. So, the halos should be a places where the dust and gases condensate. Thus in the case of very young stars one may consider the halos as the places  where planets are born. On the other hand in the case of much older stars a presence of halos should manifest itself in tiny changes of velocities and accelerations of object moving in  the considered planetary star system  relating to those calculated without taking into account of the halos presence.

If the halos are formed by unstable particles emitted by a galaxy core and these particles are such that the ratio $\frac{{\cal E}_{\phi}(t)}{{\cal E}_{\phi}^{0}}$ is suitably large inside the halos,
then rotational velocities of stars rounding the galaxy center outside the halos should differ from those calculated without taking into account the halos. Thus the halos may affect the form of rotation curves of galaxies. (Of course, we do do not assume that the   sole factor affecting  the form of the rotation curves are these halos). Another possible effect is that the
velocities of particles crossing these galactic  halos should slightly vary in time due to gravitational interactions, i. e. they should gain some acceleration. This should cause charged particles to emit electromagnetic radiation when  they cross the halo.

Note that the above mentioned effects seems to be possible to examine. All these effect are the simple consequence of the fact that the instantaneous energy ${\cal E}_{\phi_{\alpha}}(t)$ of unstable particles becomes  large  compared with ${\cal E}_{\phi_{\alpha}}^{0}$ and for some times even extremely large. On the other hand this property of ${\cal E}_{\phi_{\alpha}}(t)$ results from the rigorous analysis of properties of the quantum mechanical survival probability $a(t)$ (see (\ref{a(t)}) ) and from the assumption that the energy spectrum is bounded from below.

\end{document}